\documentclass[conference,a4paper]{IEEEtran}

\usepackage{xcolor}
\usepackage{balance}
\usepackage[normalem]{ulem}

\ifCLASSINFOpdf
  \usepackage[pdftex]{graphicx}
\else
  \usepackage[dvips]{graphicx}
\fi

\usepackage{amsmath}
\usepackage{amssymb}
\ifCLASSOPTIONcompsoc
 \usepackage[caption=false,font=normalsize,labelfont=sf,textfont=sf]{subfig}
\else
 \usepackage[caption=false,font=footnotesize]{subfig}
\fi

\usepackage{url}
\usepackage{hyperref}

\usepackage{pgf}
\usepackage{pgfplots}
\pgfplotsset{compat=1.16}

\usepackage{import}
\usepackage{siunitx}

\newcommand{\bs}{\boldsymbol}
\newcommand{\bb}{\mathbb}
\newcommand{\cl}{\mathcal}
\newcommand{\ts}{\textstyle}

\newcommand{\iid}{%
  \ifmmode
  \mathrm{i.i.d.}%
  \else%
  i.i.d.\@\xspace%
  \fi%
}

\newcommand{\ie}{\emph{i.e.}, }
\newcommand{\eg}{\emph{e.g.}, }
\newcommand{\rtpath}{\mathcal{P}}

\begin{document}
\title{Fully Differentiable Ray Tracing via Discontinuity Smoothing for Radio Network Optimization}

\author{\IEEEauthorblockN{
Jérome Eertmans,  
Laurent Jacques,  
Claude Oestges    
}                                     
\IEEEauthorblockA{
ICTEAM, Université catholique de Louvain, Louvain-la-Neuve, Belgium,
    \href{mailto:jerome.eertmans@uclouvain.be}{jerome.eertmans@uclouvain.be}}
}


\maketitle


\begin{abstract}
Recently, Differentiable Ray Tracing has been successfully applied in the field of wireless communications for learning radio materials or optimizing the transmitter orientation. However, in the frame of gradient-based optimization, obstruction of the rays by objects can cause sudden variations in the related objective functions or create entire regions where the gradient is zero. As these issues can dramatically impact convergence, this paper presents a novel Ray Tracing framework that is fully differentiable with respect to any scene parameter, but also provides a loss function continuous everywhere, thanks to specific local smoothing techniques. Previously non-continuous functions are replaced by a smoothing function, that can be exchanged with any function having similar properties. This function is also configurable via a parameter that determines how smooth the approximation should be. The present method is applied on a basic one-transmitter-multi-receiver scenario, and shows that it can successfully find the optimal solution. As a complementary resource, a 2D Python library, DiffeRT2d, is provided in Open Access, with examples and a comprehensive documentation.
\end{abstract}

\vskip0.5\baselineskip

\section{Introduction}

The field of radiocommunications has witnessed unprecedented growth over the past few decades,
largely fueled by the increase of data-intensive applications, as well as the need to access the Internet from nearly everywhere. This accessibility constraint, in addition to the other constraints faced by telecom operators, makes the task of placing antennas even more complicated. With the emergence of dynamic scenarios, such as vehicule-to-vehicule communications, it is also important to understand how to update the current configuration, \eg the orientation of the transmitting antenna, with respect to object displacements.

Traditionally, the design, analysis, and optimization of communication networks have relied on mathematical models, simulations, and measurements. Channel models are mainly either deterministic or stochastic. Based on measurements, empirical models often try to capture the general impact of the scene parameters on a given metric. For example, one could use the COST 231 Hata model \cite{costhata} to estimate the path loss between two nodes in an urban environment. This model only needs a limited amount of inputs, and considers a very general urban scenario, with no specific assumption on how the buildings are placed.
Conversely, deterministic models, like Ray Tracing (RT) or the Method of Moments, require a very precise input model, \eg the location of every building in the scene, to provide a very accurate simulation of how radio waves propagate.


While these approaches have been invaluable in the development of modern communication systems, they face limitations in simulating evolving environments, such as iteratively searching for the best antenna location or analyzing a radio link involving mobile objects like cars. Empirical models, owing to their simplicity, cannot be used for site-specific network optimization. Deterministic channel models, on the other hand, often have such a large computational cost that performing a simulation for each small variation in scene parameters could be incredibly time-consuming.

In optimization problems, it is increasingly common to utilize gradient descent methods, particularly since the emergence of automatic differentiation (AD) libraries like JAX \cite{jax2018github}. This enables differentiation of nearly every programmable function in relation to one or multiple input parameters. As such, applying AD to RT may lead to the development of a Differentiable Ray Tracing (DRT) program, hence enabling network parameter optimization.

However, the convergence of an optimizer can be dramatically impacted by the properties of the computed gradient; for instance, if it vanishes in certain regions of the parameter space, the optimization can be stuck in a suboptimal solution. In DRT, this phenomenon is directly connected to the geometry of the scene, \eg object occlusion can create discontinuous parameter transitions or zero-gradient regions. This paper proposes to address this problem by introducing a novel smoothing technique that, when applied to RT, enables to programmatically avoid those issues.

This paper is organized as follows. First, Sec.~\ref{sec:related_work} summarizes the work in DRT and the main challenges. Next, Sec.~\ref{sec:problem_definition} provides a mathematical and precise definition of the problem addressed by this work. After having motivated the use of smoothing techniques in RT in Sec.~\ref{sec:smoothing}, we present in Sec.~\ref{sec:framework} our DRT framework, published under the DiffeRT2d Python package\footnote{GitHub repository: \url{https://github.com/jeertmans/DiffeRT2d}.}. Finally, in Sec.~\ref{sec:application}, we demonstrate the capacity of our framework to improve the convergence rate of network optimization problems, before providing a few research perspectives in the concluding Sec.~\ref{sec:conclusion}.

\section{Related Work}\label{sec:related_work}

\begin{figure}[!t]
    \centering
    \import{pgf/zero_gradient}{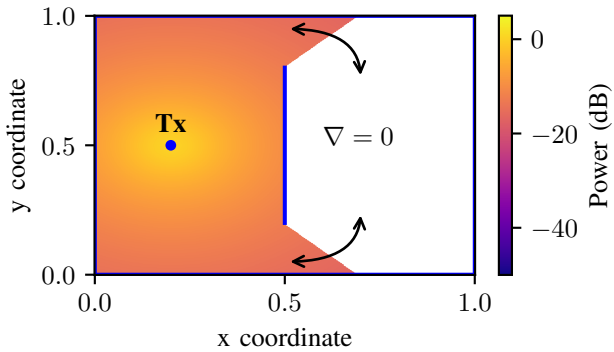}
    \caption{Illustration of gradient discontinuities, as shown by the two arrows, and a zero-gradient region (\(\nabla=0\)) in the objective function. The color gradient indicates the received power for every receiver coordinate in the scene.}
    \label{fig:problem_illustration}
\end{figure}

In computer graphics (CG), DRT is a widely studied topic, particularly in the context of reconstructing scene parameters based on images. In this field, many advances have been made, for example concerning the question of the presence of zero-gradient regions \cite{fischer2023plateau}, also referred to as \emph{plateaus}, or how to correctly carry out RT in order to take into account the abrupt transitions created by objects extremities \cite{Li:2018:DMC}.

DRT was mainly made possible thanks to AD and the development of toolboxes, such as JAX. Introduced by Wengert in 1964 \cite{wengert1964simple} and actively developed for machine learning (ML) algorithms, AD is an alternative to symbolic calculation which, conversely to finite differences, computes the true derivatives up to machine precision \cite{BARTHOLOMEWBIGGS2000171}.

While CG uses RT to perform optimization-based inverse rendering, communication operators rely on RT for channel simulation, as an alternative to actual measurements (often complicated to obtain). This difference explains why DRT is still a fairly young subject in the field of radiocommunications.

Recently, a novel tool, named Sionna \cite{hoydis2023sionna}, has been developed to provide DRT for radio propagation modeling. Sionna integrates AD within a ready-to-use RT tool that provides many features for radio channel modeling. Unfortunately, such a DRT tool can still suffer from vanishing, discontinuous gradients, mainly because RT programs partly rely on conditional statements (such as \texttt{if}, \texttt{else}), \emph{\eg} to characterize object occlusions. Fig.~\ref{fig:problem_illustration} illustrates this fact on a simple 2-D scene by showing how the change in visibility can cause discontinuities near the ends of a central wall. Moreover, all locations in the blank area receives zero contribution from the transmitting antenna, assuming only line-of-sight (LOS) path.

In the deep learning literature, the problem of vanishing gradients is well studied and has been circumvented with several strategies. For instance, the evaluation of the gradient can be artificially corrupted by noise to help the convergence in zero-gradient parameter domain \cite{neelakantan2017adding}. A similar effect is obtained in the context of reverse rendering with DRT by blurring the rendered image with a specific kernel~\cite{fischer2023plateau}. 

In this work, we propose another solution suited to radio wave propagation. We replace all discontinuous functions used in RT (\eg in conditions or occlusions) with a smoothed, parametrized function. A direct advantage of this technique is the possibility to progressively insert the objects in the scene by varying this smoothing parameter (see Sec.~\ref{sec:application}).

\section{Problem Definition}\label{sec:problem_definition}

Ray Tracing techniques approximate radio waves propagation by tracing rays between two or more communicating nodes. They can consider for instance one transmitting antenna (Tx) and one receiver antenna (Rx), bouncing rays off objects whenever an object in encountered (using reflection or diffraction models), and selecting all paths that finally reach a desired target. By computing all valid paths, geometrical optics approximations can be applied to determine, \eg the received electrical field at a given position. Because there can be many paths, it is common to limit the maximum number of allowed interactions, \eg to only simulate LOS and a limited number of reflections.

When searching for optimal parameters $\bs \gamma$ to a given problem, such as finding the best antenna position $\bs x$ to serve a fixed amount of users on multiple locations, a common approach is to use a gradient-based iterative optimizer maximizing a specific objective function \(\mathcal{F}(\bs \gamma)\), or minimizing its related loss function \(\mathcal{L}(\bs \gamma) = -\mathcal{F}(\bs \gamma)\).
A key element in this process is computing the gradient of \(\mathcal{L}\) with respect to the input parameters. Leveraging automatic-differentiation, computing this gradient becomes straightforward, as long as the objective function is differentiable.

However, limiting the number of paths in RT (\eg to reduce the computational complexity) or not simulating certain types of interactions such as diffraction impact an object visibility and may introduce discontinuities in the objective function (see Fig.~\ref{fig:problem_illustration}), \ie $\cl L$ is then only piecewise differentiable.

These discontinuities, as well as potential zero-gradient regions, present a real challenge to optimization. The next two sections explain how to build a fully DRT model and an innovative method to avoid discontinuities and zero-gradient regions. Overall, our approach improves the convergence of RT parameter optimization, as shown numerically in Sec.~\ref{sec:application}.

\section{Smoothing Local Discontinuities}\label{sec:smoothing}

The use of smoothing was first motivated by the observation that a slight inaccuracy in the position of objects in a scene can introduce a sudden variation in the number of paths linking two nodes. In urban scenarios, where the precision is often of the order of a meter \cite{osmprecision}, this poses a real problem.

An obstacle, whose frontiers can be locally modeled with a unit step function 
\begin{equation}
    \ts \theta(x) = \begin{cases} 1, &\text{if }x>0,\\ 0, &\text{otherwise,}\end{cases}
\end{equation}
\eg with \(\theta(\bs n^\top \bs x)\) in a given direction $\bs n$ when the 3-D position $\bs x$ varies, creates discontinuities in the scene. To overcome this, we replace $\theta$ with an approximating smoothed, parametrized function \(s\).

In the context of a Ray Tracing program, this function models comparisons to eliminate discontinuities. The outputs 1 and 0 of the step function, therefore of the approximation function, are associated, respectively, with \textit{true} and \textit{false} Boolean values.

\subsection{Smoothing Properties}

The approximation function, \(s\), is not unique, and anyone can craft its own function, granted that it satisfies the following properties driven by the needs of our approach.

First, the function $s$ must be parameterized with a parameter \(\alpha \in \mathbb{R}^+\) whose value increases or decreases the smoothness of the function. In particular, we impose that \(s: (x;\alpha) \in \mathbb{R} \times \bb R^+ \mapsto s(x;\alpha) \in [0;1] \) respects 
\begin{equation}
    \lim_{\alpha\rightarrow\infty} s(x;\alpha) = \theta(x).
\end{equation}

Additionally, the function must meet the following criteria:
\begin{enumerate}
    \item[{\small [C1]}] \(\lim_{x\rightarrow -\infty} s(x; \alpha) = 0\) and \(\lim_{x\rightarrow +\infty} s(x; \alpha) = 1\);
    \item[{\small [C2]}] \(s(\cdot; \alpha)\) is monotonically increasing;
    \item[{\small [C3]}] \(s(0; \alpha) = \frac{1}{2}\);
    \item[{\small [C4]}] and \(s(x; \alpha)  - s(0; \alpha) = s(0; \alpha) - s(-x; \alpha)\).
\end{enumerate}
Imposing $s\in [0,1]$ and C1 simplifies the conversion between floating-point numbers and Boolean values. The output can either be true (\(s(x; \alpha) = 1\)), false (\(s(x; \alpha) = 0\)), or fall in between these two extremes (\(0 < s(x; \alpha) < 1\)). To determine whether the output should be assigned to the true or false category, a threshold variable is used. Property C2 prevents local minima in minimization problems, while C3 and C4 are key symmetry properties for the framework presented in the next Section.

\subsection{Examples of Smoothing Functions}

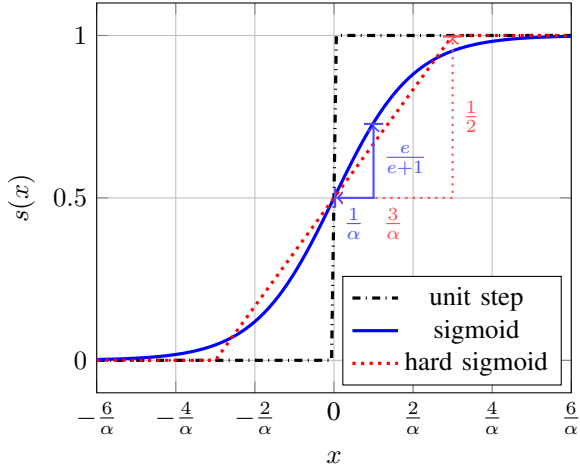
\begin{figure}[!t]
    \centering
    \begin{tikzpicture}[
    declare function={
        sigmoid(\x,\alpha)=1/(1+exp(-\x*\alpha));
        relu6(\x)=min(max(0,\x),6);
        hard_sigmoid(\x,\alpha)=relu6(\x*\alpha + 3)/6;
    }
]
\begin{axis}%
[
    grid=major,     
    xmin=-6,
    xmax=6,
    ytick={0,.5,1},
    ymin=-.1,
    ymax=1.1,
    samples=100,
    domain=-6:6,
    xlabel={\(x\)},
    ylabel={\(s(x)\)},
    height=6.75cm,
    xtick={-6,-4,-2,0,2,4,6},
    xticklabels={\(-\frac{6}{\alpha}\),\(-\frac{4}{\alpha}\),\(-\frac{2}{\alpha}\),0,\(\frac{2}{\alpha}\),\(\frac{4}{\alpha}\),\(\frac{6}{\alpha}\)},
    legend style={at={(.98,0.3)}}     
]
    \addplot[black,dash dot,very thick,mark=none]   (x,{x>0});
    \addplot[blue,very thick,mark=none]   (x,{sigmoid(x,1.0)});
    \addplot[red,very thick,dotted,mark=none]   (x,{hard_sigmoid(x,1.0)});
    \legend{unit step,sigmoid,hard sigmoid}
    
    \draw[|<->|,thick,dotted,red!70!white] (axis cs:0.0,0.5) -- (axis cs:3.0,0.5) node[below,midway] {\(\frac{3}{\alpha}\)} -- (axis cs:3.0,1.0) node[right,midway] {\(\frac{1}{2}\)};
    
    \draw[|<->|,thick,blue!70!white] (axis cs:0.0,0.5) -- (axis cs:1.0,0.5) node[below,midway] {\(\frac{1}{\alpha}\)} -- (axis cs:1.0,0.731) node[right,midway] {\(\frac{e}{e+1}\)};
\end{axis}
\end{tikzpicture}
    \caption{Various smoothing functions plotted, for a fixed \(\alpha\) parameter.}
    \label{fig:activation_functions}
\end{figure}

With some inspiration from the ML community\footnote{In ML, and in the DiffeRT2d Python package, the term \textit{activation} function is used in place of smoothing function.}, the present framework proposes two smoothing functions that satisfy all constraints C1-C4, the sigmoid and the hard sigmoid, whose \(\alpha\)-parametrization is reached by scaling
\begin{equation}
    s(x; \alpha) = s(\alpha x).
\end{equation}
The sigmoid is defined with a real-valued exponential
\begin{equation}
    \text{sigmoid}(x;\alpha) = \frac{1}{1 + e^{-\alpha x}},
\end{equation}
and the hard sigmoid is the piecewise linear function defined by
\begin{equation}
    \text{hard sigmoid}(x;\alpha) = \frac{\text{relu6}(\alpha x+3)}{6},
\end{equation}
where 
\begin{equation}
    \text{relu6}(x) = \min(\max(0,x),6).
\end{equation}

The hard sigmoid smoothing function is much cheaper to evaluate than the sigmoid, but at the cost of a non-continuous derivative at \(x=\pm3/\alpha\). Choosing an appropriate smoothing function will be dictated by the trade-off between the desired smoothing (see Fig.~\ref{fig:activation_functions}) and its related computational cost.

\section{Ray Tracing Framework}\label{sec:framework}

\begin{figure}[!t]
    \centering
    \import{pgf/power}{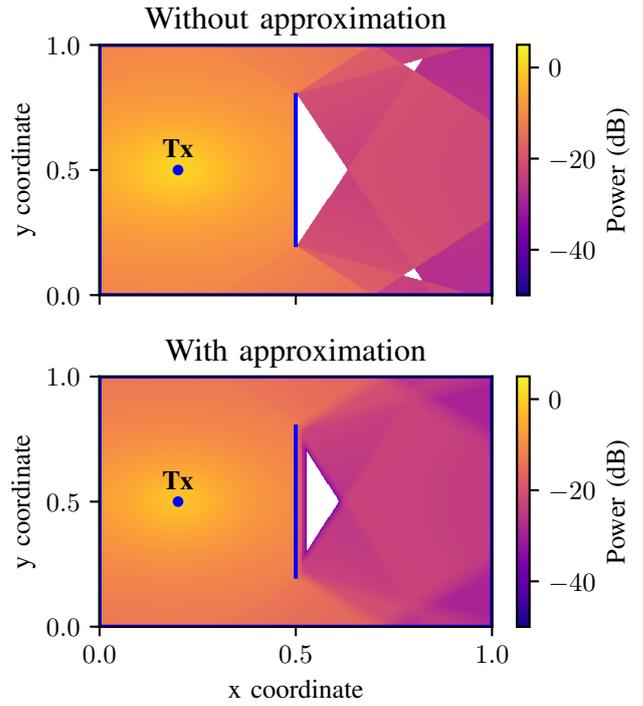}
    \caption{Power map showing the normalized received power at every coordinate in the scene. In the case of approximation (\(\alpha=50\) and \(s(x)=\text{hard\_sigmoid}(x)\)), one can observe that some zero-gradient regions have vanished, and that there is a small transition region right after the central wall. Only LOS and first order reflection paths are simulated.}
    \label{fig:power_map}
\end{figure}


Among the different possible RT implementations, the framework presented here aims at simulating the radio channel between pairs of points, the details of which are introduced in \cite{MPT}. In contrast to Ray Launching, the objective here is to exactly determine all paths up to a certain order.

Therefore, based on a given scene and a visibility matrix between all objects, an exhaustive list of all possible path candidates is created. For each of these candidates, the path can be calculated, for example using the image method (IM) or a more complete method (see Sec.~\ref{sec:path_finding}).

As discussed in \cite{MPT}, RT frameworks often separate the path tracing step into two consecutive tasks: determining path coordinates in a relaxed environment, \eg by assuming infinitely large objects, and using an appropriate path finding method (\eg IM). Then, the path is further validated, checking for objects intersection, valid reflection angles, and so on. By default, this validity check contains many discontinuities, caused by visibility changes, among others. Leveraging the mentioned approximation framework, and thus the use of an approximation function everywhere it is needed, a fully continuous path validity check can be constructed.

In a channel propagation context \cite{MPT}, the received electrical field in Rx can be computed from
\begin{equation}\label{eq:field}
    \boldsymbol{E} = \sum\limits_{\rtpath\in \mathcal{S}} V(\rtpath) \big(\,\boldsymbol{\overline{C}}(\rtpath) \cdot \boldsymbol{E}(\rtpath_1) \big),
\end{equation}
where \(\rtpath\) is one path among the set of all possible paths \(\mathcal{S}\), \(V(\rtpath)\) is a scalar value, usually 0 or 1, denoting the validity of a given path, and \(\cdot\) represents the matrix product between the dyadic coefficient, \(\boldsymbol{\overline{C}}(\rtpath)\), and the received field at the first interaction point on the path, \(\boldsymbol{E}(\rtpath_1)\). The matrix \(\boldsymbol{\overline{C}}(\rtpath)\) is the product of all Fresnel coefficients, path attenuation factors, polarization changes, and so on, from the first interaction to Rx. In LOS, \(\boldsymbol{\overline{C}}(\rtpath)\) simplifies to the identity matrix.

In \eqref{eq:field}, two components notably contain many discontinuities: the validity function, \(V(\rtpath)\), and the path itself. The following sections each discuss how to implement a continuous, but approximate, version. Fig. \ref{fig:power_map} illustrates the impact of approximation on the normalized received power.

\subsection{Continuous Validity Check}

A path can be considered invalid for many reasons: an object is blocking part of it, some coordinates are not correctly placed on the interacting objects, caused by first assuming infinitely large objects, and so on. Checking for this condition ultimately results in a Boolean, discontinuous, output.

To avoid this, the aforementioned smoothing function is applied in place of any conditional check. For example, the usual \textit{greater than} comparison, \(x_0 > x_1\), is replaced with \(s(x_0 - x_1; \alpha)\) for some $\alpha >0$. Using this, the validity check now becomes a function that returns a continuum of values between 0 and 1, where values close to 1 map to a valid path, and values close to 0 to an invalid one.

\subsection{Continuous Path Finding}\label{sec:path_finding}

Depending on the simulated types of interaction (reflection, diffraction, scattering, etc.), the desired computational time, and other criteria, a variety of path finding methods exist. The present framework implements the following three methods:

\begin{enumerate}
    \item IM, that can only handle specular reflections on planar surfaces;
    \item Fermat-Path-Tracing\footnote{FPT is a personal designation, used to indicate minimization according to Fermat's principle, as opposed to MPT which minimizes another cost function.} (FPT) \cite{FPT}, that can simulate any number of diffractions or reflections;
    \item and Min-Path-Tracing (MPT) \cite{MPT}, that can simulate any number and any type of interactions.
\end{enumerate}

All three methods find a path using an iterative approach, and have only one discontinuity: the convergence check. For this purpose, they all use the MPT loss function, as it evaluates to zero when the path is valid. Again, comparison to zero is approximated with the smoothing function \(s\). More details can be found either in the paper describing the method \cite{MPT}, or in the code provided by the DiffeRT2d toolbox.


\section{DiffeRT2d and 3D Ray Tracing} As mentioned in the introduction, the present framework is made available in open source on GitHub, and offers a self-contained Python module, DiffeRT2d, available on the Python Package Index. For demonstration purposes, the RT model is developed on two-dimensional (2D) scenes. However, it is important to note that the method is fully compatible with 3D scenes, with no assumptions made regarding the number of dimensions involved. Implementing a 3D DRT based on this framework is straightforward, although it would lengthen computation time, which is why proposing an efficient 3D implementation of this framework is a future work.

\section{Application to Antenna Location Optimization}\label{sec:application}
\begin{figure}[!t]
    \centering
    \import{pgf/optimize}{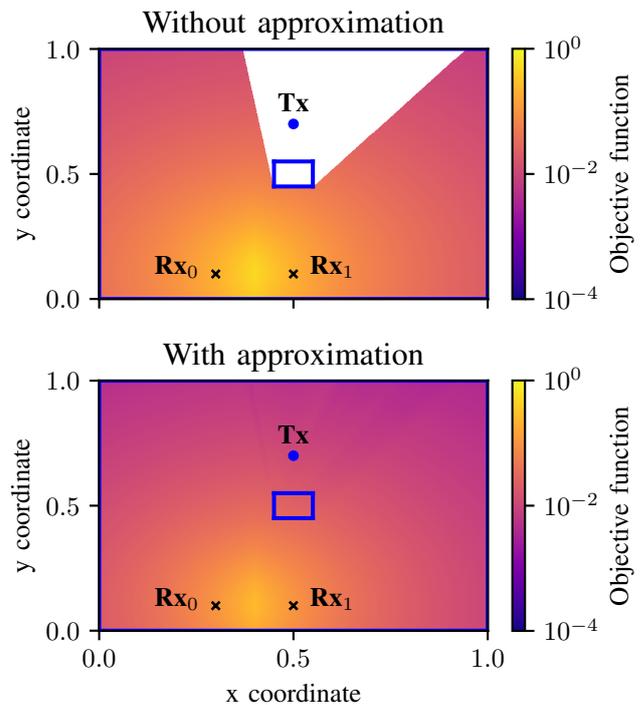}
    \caption{Illustration of the optimization problem studied, starting with \(\alpha=1\) for the approximation case. Tx is at its initialization point. For each plot, the objective function \eqref{eq:objective} has been evaluated on all scene coordinates.}
    \label{fig:opt_start}
\end{figure}
\begin{figure*}[!t]
    \centering
    \import{pgf/optimize}{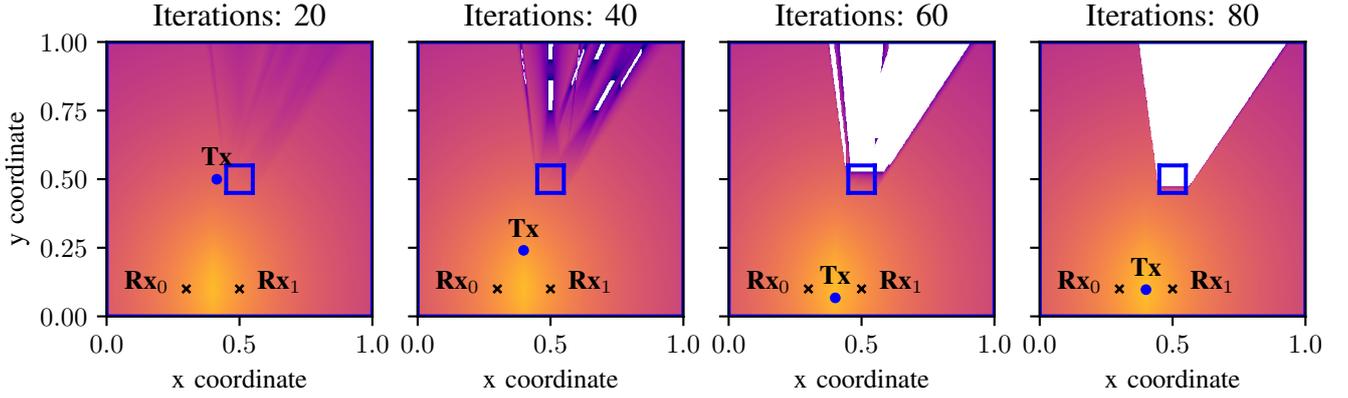}
    \caption{Illustration of the different iterations converging towards the maximum of the objective function \eqref{eq:objective}. The corresponding \(\alpha\) values are: 2.54, 6.43, 16.30 and 41.32. Here, \(\alpha\) follows a geometric progression, from 1 (see Fig.~\ref{fig:opt_start}) to 100, with a number of values equal to the number of iterations (100).}
    \label{fig:opt_steps}
\end{figure*}
In Fig.~\ref{fig:opt_start}, a basic one-transmitter-two-receiver optimization scenario is studied. To keep this example as simple as possible, only LOS paths have been simulated. The aim is to find the optimum location for the transmitting antenna, to best serve both receivers. The objective function is therefore:
\begin{equation}\label{eq:objective}
    \mathcal{F}(x, y) = \min\left(P_{\text{Rx}_0}(x, y), P_{\text{Rx}_1}(x,y)\right),
\end{equation}
where \((x,y)\) are the Tx coordinates, \(P_{\text{Rx}_i}(x, y)\) is the total power received by receiver \(i\) for a given antenna location, and the \(\min\) operator takes the minimal values between the two powers, to avoid favoring one receiver.

As shown in Fig.~\ref{fig:opt_start}, the problem has a zero-gradient region, which can cause convergence problems, especially if the starting point lies in the latter. By applying a sufficiently large smoothing, it is possible to reduce or even eliminate the zero-gradient region. This is the case when \(\alpha\) is sufficiently small, as shown in the second plot on Fig.~\ref{fig:opt_start}.

Thanks to this, it is possible to perform a gradient descent and reach the maximum of the objective function, as illustrated in Fig.~\ref{fig:opt_steps}. As this smoothing artifice is not realistic, it is important to increase the value of the \(\alpha\) parameter after each iteration, in order to obtain a value of the objective function very close to that obtained without approximation.

By repeating the optimization on random scene configurations, with two, three or four RXs, results show that using our approximation can increase the convergence success rate by 1.5 to 2. Regarding the cases where convergence was already reached without approximation, using approximating still successfully converge between 92\% and 98\% of the time.


\section{Conclusion}\label{sec:conclusion}

\balance

In this paper, a novel approach to Ray Tracing was proposed to avoid discontinuity-related issues, especially in optimization problems. By smoothing previously discontinuous conditions, the framework increases the convergence success.

The present solution offers a very customizable experience, where both the smoothing function, but also the smoothing rate, can be easily parametrized. In itself, the approximation enables to progressively introduce objects in the scene, as shown in Fig.~\ref{fig:opt_steps}. Here, all objects and functions are using the same \(\alpha\) value, but a future research direction would be to study the impact of this smoothing in every part of the code, to determine where it benefits the most. Also, the optimizer uses a very idiomatic progression for \(\alpha\) values between steps, but one could imagine a more intelligent algorithm, that only applies smoothing when the gradient is too small.

Finally, it is possible to utilize the smoothing technique to simulate semi-transparent objects, as well as refraction and attenuation through them. Specifically, the use of an intermediary \(\alpha\) value permits a path to pass through an object while simultaneously reflecting on it.

\bibliographystyle{IEEEtran}
\bibliography{IEEEabrv,biblio}

\end{document}